\documentclass[12pt]{iopart}
\usepackage{graphics}
\usepackage{rotating}
\usepackage{epsfig}
\usepackage{color}

\begin{document}

\title{Ultracold molecules: new probes on the variation of fundamental constants}
\author{Cheng Chin}
\address{James Franck Institute and Department of Physics, University of Chicago, Chicago, Illinois 60637, USA}
\author{V. V. Flambaum}
\address{School of Physics, The University of New South
Wales, Sydney NSW 2052, Australia and New Zealand Institute for
Advanced Study, Massey University, Auckland, New Zealand }
\author{M. G. Kozlov}
\address{Petersburg Nuclear Physics Institute, Gatchina
188300, Russia}

\begin{abstract}
Ultracold molecules offer brand new opportunities to probe the
variation of fundamental constants with unprecedented sensitivity.
This paper summarizes theoretical background and current constraints
on the variation of fine structure constant and electron-to-proton
mass ratio, as well as proposals and experimental efforts to measure
the variations based on ultracold molecules. In particular, we
describe two novel spectroscopic schemes on ultracold molecules
which have greatly enhanced sensitivity to fundamental constants:
resonant scattering near Feshbach resonances and spectroscopy on
close-lying energy levels of ultracold molecules.
\end{abstract}


\maketitle

\section{Introduction}\label{intro}

Creation of ultracold molecules at temperatures below 1~$\mu$K by
magneto-association \cite{Regal2003, Herbig2003} or photoassociation
\cite{Winkler2007, Danzl2008, Ni2008} opens exciting prospects to
test the variation of fundamental constants on a brand new level.
Ultracold molecules offer two major advantages: first, the very low
temperature allows the molecular samples to be fully polarized in
one single quantum state and recent experiments have developed
schemes to access any rovibrational and magnetic levels with very
high fidelity. Secondly, the success in spatially confining a large
number ($>10^4$) of ultracold molecules for a long storage time
($>$1~s) promises that an extremely high frequency resolution of
$<$1mHz$/\sqrt{{\rm Hz}}$ can be reached on molecular transitions.
These features well prepare ultracold molecules as an excellent
candidate to perform new generation, very high resolution molecular
spectroscopy.

Ultracold molecules can also be prepared in new, exotic regimes,
which offer new strategies to measure the variation of fundamental
constants. For example, Feshbach spectroscopy, which identifies
coupling to weakly-bound quantum states in atomic and molecular
collisions, can precisely determine the value of $s-$wave scattering
length \cite{Chin2008}. Here, scattering length near Feshbach
resonances can be an extremely sensitive probe on the variation of
the electron-proton mass ratio \cite{chin}. As a second example, the
variation of fundamental constants can also be strongly enhanced
when the molecules are radiatively excited to close-lying, narrow
energy levels. Several promising cases have been identified in cold
molecule systems, including Cs$_2$ \cite{DeM04} and Sr$_2$
\cite{ZKY07}.

In this review paper, we will provide an overview on the
experimental proposals of measure the variation of fundamental
constants based on ultracold molecules. As we will show below,
molecular energy structure is mostly sensitive to two fundamental
dimensionless constants: the fine-structure constant
$\alpha=\frac{e^2}{\hbar c}$, which determines the strength of the
electroweak interaction. Here $-e$ is electron's charge, $2\pi\hbar$
is Planck's constant, and $c$ is the speed of light; the second
fundamental constant is the electron-to-proton mass ratio
$\beta=m_e/m_p$, which characterizes the strength of strong
interaction in terms of the electroweak. At present, NIST lists the
following values of these constants~\cite{NIST}:
$\alpha^{-1}=137.035999679(94)$ and $\beta^{-1}=1836.15267247(80)$.

%This paper is by no mean a complete review on all ideas to test
%fundamental laws in physics with ultracold molecules. Other
%directions including test of time-reversal symmetry, and
%parity-violating interactions, see Ref.~\cite{}, can also be
%benefited by adopting ultracold molecular samples.

Remarkably, in quantum chromodynamics (QCD) there is a similar
coupling constant $\alpha_s$ for strong interaction as is $\alpha$
for electroweak interaction. However, because of the highly
nonlinear character of the strong interaction, this constant is not
well defined. Instead of $\alpha_s$, the strength of the strong
interaction is usually characterized by the parameter
$\Lambda_\mathrm{QCD}$, which has the dimension of mass and is
defined as the position of the Landau pole in the logarithm for the
strong coupling constant, $\alpha_s(r) =
C/\ln{(r\Lambda_\mathrm{QCD} /\hbar c)}$, where $C$ is a constant
and $r$ is the distance between two interacting particles.

In the Standard Model (SM), there is another fundamental parameter
with the dimension of mass -- the Higgs vacuum expectation value
(VEV), which determines the electroweak unification scale. The
electron mass $m_e$ and quark masses $m_q$ are proportional to the
Higgs VEV. Consequently, the dimensionless parameters
$X_e=m_e/\Lambda_\mathrm{QCD}$ and $X_q=m_q/\Lambda_\mathrm{QCD}$
link the electroweak unification scale with the strong scale. For
the light quarks $u$ and $d$, $X_q\ll 1$. Because of that, the
proton mass $m_p$ is proportional to $\Lambda_\mathrm{QCD}$ and
$X_e\propto \beta$. Below we will use $\beta$ instead of $X_e$
because it is more directly linked to the observables in atomic and
molecular experiments.

We begin by reviewing the theory and the state of the art of the
search for the variation of $\alpha$ and $\beta$. Next, we discuss,
in some detail, the dependence of molecular energy structure on
$\alpha$ and $\beta$. Finally, we describe possible laboratory
experiments with ultracold molecules on time-variation. Such
experiments are a novelty, and the laboratory results obtained so
far are not yet competitive. However, there are proposals for
significant improvements, and several groups have already started
implementing them.

\section{Theoretical motivation}\label{motivation}

How the variation of the physical constants and the violation of the
local position invariance may come about? Light scalar fields very
naturally appear in modern cosmological models, affecting parameters
of the SM including $\alpha$ and $\beta$ (for the whole list of SM
parameters see Ref.~\cite{Wil07}). Cosmological variations of these
scalar fields are, in turn, expected to take place because of the
drastic changes in the composition of the universe during its
evolution.

Theories unifying gravity and other interactions suggest the
possibility of spatial and temporal variation of physical
``constants'' in the universe \cite{Marciano}. Moreover, there
exists a mechanism for making all coupling constants and masses of
elementary particles both space- and time-dependent, and influenced
by the local environment (see review \cite{Uzan}). Variation of the
coupling constants could be non-monotonic, such as, e.g., damped
oscillations.

These variations are usually associated with the effect of massless
(or very light) scalar fields. One such a field is the dilaton: a
scalar which appears in string theories together with the graviton,
in a massless multiplet of closed-string excitations. Other scalars
naturally appear in those cosmological models in which our universe
is a ``brane'' floating in a space of a larger dimension. The
scalars  are simply brane coordinates in extra dimensions. However,
the only relevant scalar field recently discovered, the cosmological
dark energy, so far does not show visible variations. Observational
limits on the variation of the physical constants given in
\sref{intro} are quite stringent, allowing only for scalar
couplings, which are smaller than or, in a new class of
``chameleon-type models'', comparable to gravity \cite{OP08}.
Preliminary observations hint that there may be space variation of
$\beta$ at the level of $\delta\beta/\beta\sim10^{-8}$ \cite{LMK08}.

A possible explanation of the small observed variation was suggested
by Damour \etal\ \cite{Damour1,Damour:1994zq}, who pointed out that
cosmological evolution of scalars naturally leads to their
self-decoupling. Damour and Polyakov
%>mgk [ref]
\cite{Damour:1994zq}
%<mgk
have further suggested that the variations should take place when
the scalars become excited by some physical change in the universe,
such as phase transitions, or by other drastic changes in the
equation of state of the universe. They considered several phase
transitions, but since the publication of their paper a new
transition has been discovered, one from a matter dominated
(decelerating) era to a dark-energy dominated (accelerating) era.
This transition is a relatively recent event, corresponding to a
cosmological redshift $z\approx 0.5$, or a look-back time of
approximately 5 billion years.

The time dependence of the perturbation related to the transition
from the decelearting to the accelerating era could be calculated
\cite{Barrow,Olive}. The calculation shows that the self-decoupling
process is effective enough to explain why \emph{after} this
transition the variation of the constants is as small as observed in
the present-time laboratory experiments. However, the calculated
time dependence is also consistent with the observations of the
variation of the proton-to-electron mass ratio and the
electromagnetic fine-structure constant at $z\geq 1$
\cite{Murphy,Ubach,LML07}.

\section{Current constraints on the variation of $\alpha$ and $\beta$}\label{intro}
The analysis of the data from the Big Bang nucleosynthesis
\cite{Dmitriev}, quasar absorption spectra, and the Oklo natural
nuclear reactor yields the space-time variation of the constants on
the time scale of the lifetime of the universe, i.e. from a few
billion to more than ten billion years. In comparison, the
frequencies of different atomic and molecular transitions in the
laboratory experiments yield a time variation on the timescale from
a few months to a few years. There is no model-independent
connection between the variations on such different timescales.
However, in order to compare the astrophysical and laboratory
results, we often assume a linear time dependence of the constants.
In this way, we can interpret all the results in terms of the time
derivatives of the fundamental constants. Within this assumption, we
can use the quasar absorption spectra to obtain the best current
limit on the variation of the mass ratio $\beta$ and of $X_e$,
\cite{FK1,MM08}

 \begin{equation}\label{best_mu_dot}
 \dot{\beta}/\beta=\dot{X_e}/X_e=(-1.2 \pm 1.4) \times
 10^{-16}\mathrm{~yr}^{-1}\,.
 \end{equation}

Combining this result with the atomic clock results \cite{clock_1},
one can obtain upper bounds on the variation of $\alpha$
\cite{tedesco,FS2007,Fla07}. The best upper bound on the variation
of $\alpha$ is provided by the single-ion optical clock experiments
as \cite{R08}

 \begin{equation}\label{best_alpha_dot}
 \dot{\alpha}/\alpha=(-1.6 \pm 2.3) \times 10^{-17}\mathrm{~yr}^{-1}\,.
 \end{equation}

The measurements at the Oklo natural reactor provide the best bound
on the variation of $X_s=m_s/\Lambda_\mathrm{QCD}$, where  $m_s$ is
the strange quark mass \cite{Shl76,Oklo,FlambaumShuryak2002},

 \begin{equation}\label{best_X_s_dot}
 |\dot{X_s}/X_s| < 10^{-18}\mathrm{~yr}^{-1}\,.
 \end{equation}

Note that the Oklo data cannot yield any bound on the variation of
$\alpha$ since the effect of $\alpha$ there is much smaller than the
effect of $X_s$ and should be neglected within the accuracy of the
present theory \cite{FlambaumShuryak2002}.

\section{Dependence of atomic and molecular spectra on $\alpha$ and $\beta$}

Atomic and molecular spectra are most naturally described in atomic
units $(\hbar=m_e=e=1)$, with the energy measured in Hartrees (1
Hartree = $\frac{e^4 m_e}{\hbar^2}$ = 2~Rydberg =
219474.6313705(15)~cm$^{-1}$). In these units, the nonrelativistic
Schr\"{o}dinger equation for an atom with an infinitely heavy
pointlike nucleus does not include any dimensional parameters. The
dependence of the spectrum on $\alpha$ appears only through
relativistic corrections, which describe the fine-structure, the
Lamb shift, etc. The dependence of atomic energies on $\beta$ is
known as the isotope effect and is caused by a finite nuclear mass
and volume. There are even smaller corrections to atomic energies,
which depend on both $\alpha$ and $\beta$ and are known as the
hyperfine structure.

One could argue that the atomic energy unit itself depends on $\alpha$ as
it can be expressed as $\alpha^2m_e c^2$, with $m_e c^2$ the
rest energy of a free electron. However, an experimental search for
a possible variation of the fundamental constants relies on the observation of the time-variations of the ratios of different transition frequencies to
one another. In such ratios, the dependence of the units on the
fundamental constants cancels out. Below we will use atomic units
unless stated otherwise.

The relativistic corrections to the binding energies of the atomic valence
electrons are on the order of $\alpha^2 Z^2$, where $Z$ is the atomic
number, and become quite large for heavy elements. For our purposes,
it is convenient to consider the dependence of the atomic transition
frequencies on $\alpha^2$ in the form
 \begin{equation}\label{K-factor}
 \omega = \omega_0( 1 + \frac{K}2 x),
 \end{equation}
where $x = (\frac{\alpha}{\alpha_0})^2 - 1 \approx \frac{2 \delta
\alpha}{\alpha}$ and $\omega_0 $ is a transition frequency for
$\alpha=\alpha_0$. Rough estimates of the enhancement factor
$K=\delta(\log \omega)/\delta(\log \alpha)$ can be obtained from
simple one-particle models, but in order to obtain accurate values
one has to account for electronic correlations via large-scale
numerical calculations. Recently, such calculations have been
carried out for many atoms and ions
\cite{dzuba1999,Dy,q,nevsky,BEIK06,PKT07,DJ07,DF07b}.

Isotope effects in atoms are on the order of $\beta\sim 10^{-3}$ and
the magnetic hyperfine structure roughly scales as $\alpha^2 \beta Z
g_\mathrm{nuc} \sim 10^{-7}Zg_\mathrm{nuc}$, where $g_\mathrm{nuc}$
is nuclear $g$-factor. One has to keep in mind that $g_\mathrm{nuc}$
depends on the quark parameters $X_q$. This dependence has to be
considered when comparing, e.g., the frequency of the hyperfine
transition in $^{133}$Cs (Cs frequency standard) \cite{tedesco} or
the hydrogen 21~cm hyperfine line \cite{TWM05,TWM07} with various
optical transitions \cite{tedesco}. At present there are many very
accurate experiments comparing different optical and microwave
atomic clocks
\cite{clock_1,prestage,Marion2003,Bize2005,Peik2004,Bize2003,
Fischer2004,Peik2005,Peik2006,R08}. A detailed discussion of the
atomic experiments can be found in recent reviews
\cite{karshenboim,Lea07}.

Molecular spectroscopy opens additional possibilities to study the
variation of fundamental constants. It is known that $\beta$ defines
the scales of electronic, vibrational, and rotational intervals in
molecular spectra, $E_\mathrm{el} : E_\mathrm{vib} : E_\mathrm{rot}
\sim 1 : \beta^{1/2} : \beta$. In addition, molecules have fine and
hyperfine structure, $\Lambda$-doubling, hindered rotation, etc. All
these effects have different dependencies on the fundamental
constants.

The sensitivity to temporal variation of the fundamental constants
may be strongly enhanced in coupling between molecular levels and
continuum, as well as transitions between narrow close-lying levels
of different types.

In the following, we describe selected cases to illustrate the
potential of ultracold molecules in probing the constants:
scattering lengths of ultracold atomic and molecular collisions,
narrow close-lying levels of diatomic molecules, and the proposed
experiment with ultracold Cs$_2$ and Sr$_2$ molecules. Enhancement
of the relative variation $\delta\omega/\omega$ can also occur in
transitions between nearly degenerate levels of atoms
\cite{dzuba1999,Dy,nevsky,budker,budker1}, thermal molecules
\cite{DeM04,mol,VKB04,FK1,FK2}, and nuclei \cite{th,th4}.

\section{Enhanced sensitivity of scattering length on $\beta$ in
ultracold atomic and molecular collisions near Feshbach resonances }

An interesting case of the enhancement of the effect of the
variation of fundamental constants arises in collisions of ultracold
atoms and molecules near Feshbach resonances \cite{chin}. In cold
collisions, scattering phase shift $\phi$ is very sensitive to the
change of the electron-proton mass ratio $\beta$. This dependence
can be understood as $\phi$ depends on the ratio of the molecular
potential $V$ and the atomic mass $m_{at}$, namely,
$\phi\propto(V/m_{at})^{1/2}$. Since the molecular potential $V$ is
predominately electronic and the atomic mass is essentially
baryonic, we have
$\phi\sim(V/m_{at})^{1/2}\sim(m_e/m_p)^{1/2}\sim\beta^{1/2}$. A
model potential calculation shows that, among other cold collisions
properties, scattering length $A$ can be extremely sensitive to
$\beta$. A fractional variation of $\beta$ can be amplified to a
change of $A$ according to:

 \begin{equation}\label{d_a_final}
 \frac{\delta A}{A}=K\frac{\delta\beta}{\beta}\,,
 \end{equation}

\noindent where the enhancement factor $K=\delta(\log A)/\delta
(\log \beta)$ can be as large as $10^9\sim10^{12}$ near narrow or
optical Feshbach resonances \cite{chin}. Currently the best
measurement on the scattering length has an uncertainty of
$10^{-4}$, which can potentially probe the variation of $\beta$ on
the level of $10^{-13}\sim10^{-16}$.

Recently, Gibble group at Penn State University demonstrated that
atomic-clock type experiment can determine the scattering phase
shift with very high precision \cite{Hart2007}. They concluded that
1-p.p.m. ($10^{-6}$) measurement on the scattering length can be
reached \cite{Hart2007}. If such a measurement is performed near
narrow Feshbach resonances in two consecutive years, variation of
$\beta$ can be probed on the level of $10^{-15}\sim10^{-18}$ /yr.

Note that the calculation of the factor $K$ in Ref.~\cite{chin} is
based on the analytic formula for the scattering length derived in
Ref.~\cite{Gribakin}. This formula is valid for an arbitrary
interatomic potential with an inverse-power long-range tail
($-C_6/r^6$ for neutral atoms, where $C_6$ is the van der Waals
potential constant and $r$ is the atomic separation), i.e., it
includes all anharmonic corrections.

To the best of our knowledge, it is the only suggested experiment on
time-variation where the observable is not a frequency. However,
another parameter, $L$, with the dimension of length is needed to
compare $A$ with and thus render it dimensionless. In Ref.~
\cite{chin} the scattering length was defined in atomic units
$(a_B)$. It is important, however, that because of the large
enhancement in \Eref{d_a_final}, the possible dependence of $L$ on
$\beta$ becomes irrelevant. For example, if we measure $A$ in
conventional units, meters, which are linked to the Cs standard,
then $\delta L/L =-\delta\beta/\beta$, and
 \begin{equation}\label{d_aL}
 \frac{\delta (A/L)}{(A/L)}=(K+1)\frac{\delta\beta}{\beta}\,.
 \end{equation}
As long as $K\gg 1$, the dependence of the units used on the
fundamental constants can be neglected. Below, we discuss several
other experiments with huge enhancement factors, where this argument
can also be applied.

\section{Narrow close-lying levels of diatomic molecules}\label{diatomics}

In this section we focus on narrow close-lying levels of varying
nature in diatomic molecules. Such levels may occur due to the
cancelation between either hyperfine and rotational structures
\cite{mol}, or between the fine and vibrational structures within the
electronic ground state \cite{FK2}. The intervals between the levels
correspond to microwave frequencies, which are experimentally accessible, and have narrow linewidths, typically $\sim 10^{-2}$~Hz.
The enhancement of the relative variation $K$ can exceed $10^5$ in such cases.

\subsection{Molecules with cancelation between hyperfine
structure and rotational intervals} \label{hfs-rot}

Consider diatomic molecule with the $^2\Sigma$ ground state (one
unpaired electron). Examples of such molecules include LaS, LaO,
LuS, LuO, and YbF \cite{HH79}. The hyperfine interval
$\Delta_\mathrm{hfs}$ is proportional to $\alpha^2 Z
F_\mathrm{rel}(\alpha Z) \beta g_\mathrm{nuc}$, where
$F_\mathrm{rel}$ is an additional relativistic (Casimir) factor
\cite{Sob79}. The rotational interval $\Delta_\mathrm{rot} \propto
\beta$ is approximately independent of $\alpha$. If we find a
molecule with $\Delta_\mathrm{hfs} \approx \Delta_\mathrm{rot}$, the
splitting $\omega$ between hyperfine and rotational levels will
depend on the following combination
 \begin{equation}
 \label{hfs-rot1}
 \omega \propto \beta \left[\alpha^2 F_\mathrm{rel}(\alpha Z)\, g_\mathrm{nuc}
 - \mathrm{const}\right]\, .
 \end{equation}
The relative variation is then given by
\begin{equation}
\label{hfs-rot2}
 \frac{\delta\omega}{\omega}
 \approx \frac{\Delta_\mathrm{hfs}}{\omega}
 \left[\left(2+K\right)\frac{\delta\alpha}{\alpha} + \frac{\delta
 g_\mathrm{nuc}}{g_\mathrm{nuc}}\right]+\frac{\delta\beta}{\beta}\,,
\end{equation}
where the factor $K$ comes from the variation of $F_\mathrm{rel}(\alpha
Z)$, and for $Z \sim 50$, $K\approx 1$. As long as
$\Delta_\mathrm{hfs}/\omega\gg 1$, we can neglect the last term in
\Eref{hfs-rot2}.

The data on the hyperfine structure of diatomics are hard to come by and usually
not very accurate. Using the data from
\cite{HH79}, one can find that $\omega = (0.002\pm 0.01)$~cm$^{-1}$
for ${}^{139}$La${}^{32}$S \cite{mol}. Note that for $\omega =
0.002$~cm$^{-1}$, the relative frequency shift is
\begin{equation}
\label{hfs-rot3}
 \frac{\delta\omega}{\omega}
 \approx 600\,\frac{\delta\alpha}{\alpha}\,.
\end{equation}
As new data on molecular hyperfine constants become available,
it is likely that other molecular candidates with the cancelation effect
will be found.

\subsection{Molecules with cancelation between fine-structure
and vibrational intervals} \label{fs-vib}

The fine-structure interval, $\omega_f$, increases rapidly with the
nuclear charge Z:
\begin{equation}
\label{of}
 \omega_f \sim Z^2 \alpha^2\, ,
\end{equation}
In contrast, the vibrational energy quantum decreases with the
atomic mass
\begin{equation}
 \label{ov}
\omega_\mathrm{vib} \sim M_r^{-1/2} \beta^{1/2}\, ,
\end{equation}
where the reduced mass for the molecular vibration is $M_r m_p$.
Therefore, we obtain an equation $Z=Z(M_r,v)$ for the spectral lines at fixed $Z,M_r$ where we can expect an approximate cancelation between
the fine-structure and vibrational intervals:
\begin{equation}
 \label{o}
 \omega=\omega_f - v\,  \omega_\mathrm{vib} \approx 0 \,,
 \quad v=1,2,...
\end{equation}
Using Eqs.~(\ref{of}--\ref{o}), it is easy to find the dependence of
the transition frequency on the fundamental constants
\begin{equation}
 \label{do}
 \frac{\delta\omega}{\omega}=
 \frac{1}{\omega}\left(2 \omega_f \frac{\delta\alpha}{\alpha}+
 \frac{v}{2} \omega_\mathrm{vib} \frac{\delta\beta}{\beta}\right)
 \approx K \left(2 \frac{\delta\alpha}{\alpha}+
\frac{1}{2} \frac{\delta\beta}{\beta}\right),
\end{equation}
where the enhancement factor, $K= \frac{\omega_f}{\omega}$,
%>mgk11/07 and 29/08>
is due to the relative frequency shift for a given change of the
fundamental constants. Large values of the factor $K$ are
experimentally favorable, as the correspondingly large relative
shifts can be more easily detected. However, large value of $K$ do
not always guarantee a more sensitive measurement. In some cases of
quasi-degenerate levels, this factor may become irrelevant
\cite{budker}. Thus, it is also important to consider the absolute
values of the shifts and compare them with the linewidths of the
transitions in question.

Assuming $\delta \alpha / \alpha \sim 10^{-17}$ and $\omega_f\sim
500$~cm$^{-1}$, we obtain $\delta\omega \sim 10^{-14}$ cm$^{-1}\sim
3 \times 10^{-4}$ Hz. In order to obtain a similar sensitivity from
a comparison of the hyperfine transition frequencies of Cs or Rb,
one would have to measure the shifts with an accuracy of $\sim
10^{-7}$ Hz. Note that discussed here narrow close levels exist, for
example, in the molecular ion Cl$_2^+$ and in the molecule SiBr
\cite{VKB04}.

\section{Proposed experiments with C\lowercase{s}$_2$ and S\lowercase{r}$_2$}\label{Cs2}

In this section we discuss two recently proposed experiments with
cold diatomic molecules, one with Cs$_2$ at Yale \cite{DeM04,DSS07}
and one with Sr$_2$ at JILA \cite{ZKY07}.

% \begin{figure*}[htb]
%  \includegraphics[scale=1.0]{Fig_Cs2}
%  \caption{Levels $^3\Sigma_u^+$ and $^1\Sigma_g^+$ in Cs$_2$ molecule (figure
%  from Ref.~\cite{Sai05}).} \label{fig_Cs2}
% \end{figure*}

The Yale experiment is based on the idea described in
Ref.~\cite{DeM04} to match the electronic energy with a large number
of vibrational quanta. The difference compared with
Eqs.~(\ref{of}~--~\ref{o}) is that here the electronic transition is
between a  $^1\Sigma_g^+$ ground state and a $^3\Sigma_u^+$ excited
state and thus, to the first approximation, its frequency is
independent of $\alpha$. The energy of this transition is about
3300~cm$^{-1}$ and the number of the vibrational quanta needed to
match it is on the order of 100. For the vibrational quantum number
$v \sim 100$, the density of the levels is high due to the
anharmonicity of the potential and hence it is possible to find two
nearby levels belonging to two different potential energy curves.
This leads to an enhanced sensitivity to variation of $\beta$, as in
\Eref{o}. Cold Cs$_2$ molecules in a particular quantum state can be
produced by the photoassociation of cold Cs atoms in a trap.

Let us estimate the sensitivity of this proposed experiment to the
variation of $\alpha$ and $\beta$. If we neglect the anharmonicity,
we can write the transition frequency between the closely-spaced
vibrational levels of the two electronic terms as
\begin{equation}\label{Cs2a}
 \omega=\omega_\mathrm{el,0}+qx
 +(v_2+\frac12)\,\omega_\mathrm{vib,2}
 -(v_1+\frac12)\,\omega_\mathrm{vib,1},
\end{equation}
where $v_2\ll v_1$. The dependence of this frequency on $\alpha$ and
$\beta$ is given by
\begin{equation}\label{Cs2b}
 \delta\omega\approx 2q\frac{\delta\alpha}{\alpha}
 -\frac{\omega_\mathrm{el,0}}{2}\frac{\delta\beta}{\beta}\,,
\end{equation}
where we made use of the inequality $\omega\ll\omega_\mathrm{el,0}$.
For the ground state of atomic Cs, the $q$-factor is about
1100~cm$^{-1}$, which is close to $\frac14 \alpha^2 Z^2 E_{6s}$,
where $E_{6s}$ is the ground-state binding energy. If we assume that
the same relation holds for the electronic transition in the
molecule, we obtain $|q|\sim\frac14 \alpha^2
Z^2\omega_\mathrm{el,0}\sim 120$ cm$^{-1}$. Using this rough
estimate and \Eref{Cs2b} we have (in cm$^{-1}$)
\begin{equation}\label{Cs2c}
 \delta\omega\approx -240\frac{\delta\alpha}{\alpha}
 -1600\frac{\delta\beta}{\beta}\,,
\end{equation}
where we assumed that the relativistic corrections reduce the
dissociation energy of the molecule, as a result of which $q$ is
negative. This estimate shows that the experiment with Cs$_2$ is
mostly sensitive to the variation of $\beta$.

As noted above, for high vibrational states, the potential is highly
anharmonic. This significantly decreases the sensitivity as
estimated by Eq.~(\ref{Cs2c}). This can be seen either from the WKB
approximation \cite{DeM04,DSS07}, or from an analytic solution for
the Morse potential \cite{ZKY07}. The quantization condition for the
vibrational spectrum in the WKB approximation
\begin{equation}\label{Cs2d}
 \int_{R_1}^{R_2}\sqrt{2M(U(r)-E_n)}\,\mathrm{d}r
 = \left(v+\frac12\right)\pi\,.
\end{equation}
yields, by differentiation with respect to  $\beta$, the following
result
\begin{equation}\label{Cs2e}
 \delta E_v =\frac{v+\frac12}{2\rho(E_v)} \frac{\delta\beta}{\beta}\,,
\end{equation}
where $\rho(E_v)\equiv\left(\partial E_v/\partial
v\right)^{-1}\approx \left(E_v-E_{v-1}\right)^{-1}$ is the level
density. For the harmonic part of the potential,
$\rho=\mathrm{const}$ and the shift $\delta E_v$ increases linearly
with $v$, but for vibrational states near the dissociation limit,
the level density $\rho(E) \longrightarrow \infty$ and $\delta E_v
\longrightarrow 0$. Consequently, the maximum sensitivity $\sim
1000$ cm$^{-1}$ is reached at $v\approx 60$, and rapidly drops for
higher $v$. The Yale group has found a conveniently close
vibrational level of the upper $^3\Sigma_u$ state with $v=138$. The
sensitivity for this level is however only $\sim 200$ cm$^{-1}$
\cite{DSS07}. There are still good prospects for finding other
close-lying levels with smaller $v$, for which the sensitivity may
be several times higher.

The sensitivity as given by Eq.~(\ref{Cs2c}) towards the variation
of $\alpha$ is also reduced by the anharmonicity of the potential.
For the highest vibrational levels of the electronic ground state as
well as for all levels of the upper (weakly bound) electronic state,
the separation between the nuclei is large, $R\approx 12$~a.u. Thus,
both
%>mgk electronic
%<mgk
electronic wave functions are close to either symmetric (for
$^1\Sigma_g^+$) or antisymmetric combination (for $^3\Sigma_u^+$) of the
atomic $6s$ functions,
\begin{equation}
 \label{Cs2wf}
 \Psi_{g,u}(r_1,r_2) \approx\frac{1}{\sqrt{2}}
 \left(6s^a(r_1) 6s^b(r_2) \pm 6s^b(r_1) 6s^a(r_2)\right).
\end{equation}
As a result, all the relativistic corrections are (almost) the same for
both electronic states.

The deleterious effect of the anharmonicity on the sensitivity to
the variation of $\beta$ and $\alpha$ can be also obtained from the
analysis of the Morse potential. Its eigenvalues are given by
\begin{equation}
\label{Cs2g}
E_v=\omega_0(v+\frac12)-\frac{\omega_0^2(v+\frac12)^2}{4d}-d\,,
\end{equation}
with $\omega_0=2\pi a \sqrt{2d/M}$ and $d$ the dissociation energy.
The last eigenvalue $E_N$ is found from the conditions $E_{N+1}\le
E_N$ and $E_{N-1}\le E_N$. Clearly, $E_N$ is very close to zero and
is independent of $\beta$ and $\alpha$ and thus of their variation.

% \begin{figure*}[htb]
%  \includegraphics[scale=0.8]{RamanScheme}
%  \caption{The scheme for Raman spectroscopy of Sr$_2$ ground-state
%  vibrational spacings. A two-color photoassociation pulse prepares
%  molecules in the $v=v_\mathrm{max}-2$ vibrational state (denoted on the plot
%  as $v=-3$). Subsequently, a Raman pulse couples the $v=-3$ and $v=27$ states via
%  $v'\approx 40$ level of the excited $0_u^+$ state (figure
%  from Ref.~\cite{ZKY07}).} \label{fig_Sr2}
% \end{figure*}

We note that the highest absolute sensitivity can be expected for
vibrational levels in the middle of the potential energy curve.
However, in this part of the spectrum, there are no close-lying
levels of a different state that would allow to maximize the
relative sensitivity $\delta\omega/\omega$. Zelevinsky \etal\
\cite{ZKY07} proposed to measure molecular transition with the
highest sensitivity to $\beta$ using a frequency comb. In the
proposed experiment, an optical lattice is used to trap Sr$_2$
molecules formed in one of the uppermost vibrational levels of the
ground electronic states by photoassociation. In the next step, a
Raman transition is proposed to create molecules in one of the most
sensitive levels in the middle of the potential well. To minimize
light shifts on the relevant transition, the optical lattice is
formed by a laser operated at the ``magic'' wavelength \cite{ZKY07}.
Sensitivity to the external magnetic field can also be greatly
reduced by choosing molecular levels with identical magnetic moment.

The frequency comb scheme offers the highest possible absolute
sensitivity for a given molecule. Unfortunately, the dissociation
energy of Sr$_2$ is only about 1000~cm$^{-1}$, which is 3 times
smaller than that of Cs$_2$. Consequently, the highest sensitivity
for the Sr$_2$ molecule occurs at about 270~cm$^{-1}$, i.e. only
slightly higher than for the $v=138$ level in Cs$_2$. Therefore, it
may be useful to try to apply this scheme to another molecule with a
larger dissociation energy. Finally, we note that the sensitivity to
$\alpha$-variation in the Sr$_2$ experiment is additionally reduced
by a factor $(55/38)^2\approx 2$ because of the smaller $Z$.

\section{Conclusions}\label{concl}

Astrophysical observations of the spectra of diatomic and polyatomic
molecules can reveal a possible variation of the electron-to-proton
mass ratio $\beta$ on a time scale from 6 to 12 billion years.
However, the astrophysical results obtained so far are inconclusive.
Much of the same can be said about the astrophysical search for an
$\alpha$-variation.

The use of cold molecules holds the promise to dramatically enhance
the sensitivity of the molecular experiments. An example is the
scattering length in cold collisions of atoms and molecules near
narrow Feshbach resonances, which can have a surprisingly high
sensitivity on the variation of $\beta$ \cite{chin}. Measurement of
scattering phase shift in atomic clock-type experiments
\cite{Hart2007} can potentially test the fractional variation of
$\beta$ on the level of 10$^{-15}\sim10^{-18}$/yr.

Preliminary spectroscopic experiment with ultracold Cs$_2$ molecules
have been recently performed at Yale \cite{DSS07}. The electronic
transition between the $^3\Sigma_u^+$ and $^1\Sigma_g^-$ states of
Cs$_2$ is independent, to the first approximation, of $\alpha$. On
the other hand, the sensitivity to the $\beta$-variation may be
enhanced because of the large number of the vibrational quanta
needed to match the electronic transition. However, the
anharmonicity of the potential suppresses this enhancement for very
high vibrational levels near the dissociation limit. As a result,
the sensitivity to the variation of $\beta$ for the $v=138$ level is
about the same as that given in Eq.~(13). It is possible that there
are other close-lying levels with smaller vibrational quantum number
and, consequently, allow for a higher sensitivity. Even if such
levels are not found, the experiment with the $v=138$ level may
improve the current limit on the time-variation of $\beta$ by
several orders of magnitude.

An experiment with the Sr$_2$ molecule has been recently proposed at
JILA \cite{ZKY07}. This experiment has a similar sensitivity to the
time-variation of $\beta$ as the experiment with Cs$_2$; these
experiments are complementary to the experiments with the molecular
radicals, which are mostly sensitive to the time-variation of
$\alpha$ \cite{FK2}.

\section{Acknowledgments}
We want to thank J.\ Ye, D.\ DeMille, and S.\ Schiller for their
extremely useful comments. C.C.\ acknowledges support from the
NSF-MRSEC program under No. DMR-0820054 and ARO Grant No.
W911NF0710576 from the DARPA OLE Program. V.F.\ acknowledges support
from Masden grant and Australian Research Council. M.K.\
acknowledges support from RFBR grant 08-02-00460.
\\
%\bibliographystyle{apsrev}%
%\bibliography{../bib/alpha,../bib/julia_w,../bib/my_ref_w}
%\end{document}

%%%%  >>>>eoref<<<<<
%%%%  >>>>eoref<<<<<
%%%%  >>>>eoref<<<<<
%%%%  >>>>eoref<<<<<

\end{document}